\documentstyle[12pt]{article}
\newcommand{\be}{\begin{equation}}
\newcommand{\ee}{\end{equation}}
\newcommand{\ben}{\begin{enumerate}}
\newcommand{\een}{\end{enumerate}}

\newcommand{\dis}{\displaystyle}

\newcommand{\Ph}{\mbox{$\bf Ph$}\, }

\font \msb=msbm10 scaled \magstep1

\newcommand{\bR}{\mbox{\msb R} }

\font \eul=eufm10 scaled \magstep2

\newcommand{\gotK}{\mbox{\eul k}}

\begin{document}

\title{\bf On the classical $\kappa $-particle}
\author{{\bf S. Zakrzewski}  \\
\small{Department of Mathematical Methods in Physics,
University of Warsaw} \\ \small{Ho\.{z}a 74, 00-682 Warsaw, Poland} }

\date{}
\maketitle

\begin{abstract}
The dependence of velocity on momentum for the free massive
particle obeying the $\kappa $-Poincar\'{e} Poisson symmetry is
calculated in terms of intrinsic non-commuting space-time
coordinates and shown to have a monotonic character, with upper
limit of velocity equal to 1.
\end{abstract}

\section*{Introduction}

The theory of quantum groups \cite{D} (and Poisson groups
\cite{D:ham,S-T-S,Lu-We,Lu}) provides us
with `non-commutative' models of space-time which (depending on
some deformation parameter) can be arbitrarily close to the
usual `commutative' model. In this situation it is rather
reasonable to study how the known models of physical systems
behave when going to the non-commutative case.

An idea how to construct a model of a classical (free) particle
`moving' in a Poisson Minkowski space-time $M$ has been given in
\cite{poi} (see also \cite{poican}). The natural geometric way
is to replace the cotangent bundle to $M$ (the {\em  extended phase
space} with coordinates $q^k,p_l$, $k,l=0,1,2,3$) by the
symplectic groupoid of the Poisson manifold $M$ (called also
the {\em twisted cotangent bundle} of $M$, or the {\em phase
space} of $M$, see \cite{poi,poican,srni,abel} and references
therein), which we denote by $\Ph M$. Then the Poisson
action of the underlying Poisson Poincar\'{e} group $G$ on $M$
yields so called {\em moment map} from $\Ph M$ to the dual
Poisson group $G^*$ of $G$. Taking the inverse image of a
symplectic leaf (an orbit of the dressing action) in $G^*$ by
the moment map, we obtain the {\em mass shell} in this new
context.

The characteristics on the mass shell (given by the directions
of degeneracy of the symplectic form on the shell) are
interpreted as phase trajectories of the particle. We can
project these characteristics down to $M$, even if the
interpretation of these `world lines' is not quite clear. Note
that there are two projections in the `twisted cotangent bundle'
and one can hesitate which to use. One should probably choose
always the left projection (recall that it is the left
projection which reproduces the Poisson structure on $M$: the
Poisson bracket of functions on $M$ coincides with their bracket
in $\Ph M$ when pulled back by the left projection) and to admit
that each model has its `mirror image'. Anyway, these `world
lines' exist in the model.

In \cite{poi} we have calculated phase trajectories and world
lines in the $D=2$ case. In \cite{poican} we have considered in
the same model another set of position variables (another
projection in $\Ph M$). These variables, in contrast to the
previous ones, were Poisson commuting, although introduced
somewhat `by hands'. The world lines obtained using these
non-intrinsic `commuting positions' seemed to be even more
`realistic' than the previous ones.

An opposite situation seems to happen in the model based on the
Poisson Poincar\'{e} group \cite{luki} corresponding to the
$\kappa $-deformation \cite{Lukier,Lukier1} in $D=4$.  In the
present paper we calculate the world lines in terms of
non-commuting original positions for this model.  We show that
the dependence of the velocity on the momentum is monotonic
(with the upper limit $= 1$), in contrast with the unpleasant
behavior observed recently in \cite{woj}, when using commuting
positions which appear also in this model.

\section{The twisted cotangent bundle to $\kappa $-Minkowski}

Let $G$ denote the Poincar\'{e} group. The Poisson structure on
$G$ corresponding to the $\kappa $-deformation is fixed by the
classical $r$-matrix
$$ r= h(L_1\wedge P_1 + L_2\wedge P_2 + L_3\wedge
P_3),\qquad\qquad h\equiv \frac{1}{\kappa} $$
(cf. \cite{luki}) on the Poincar\'{e} Lie algebra, where $L_i,
P_j$ denote the `boost' generators and the spatial translations
generators, respectively. The corresponding Poisson Minkowski
space is the usual Minkowski space $M$ (which we may identify
with the subgroup $V\subset G$ of translations), with the
following Poisson structure
\be\label{pb}
 \{ q^k ,q^0\} = hq^k,\qquad\qquad \{q^k,q^l\} =0,\qquad\qquad
k,l=1,2,3.
\ee
Due to the linearity of the Poisson bracket, $M$ is the dual
space of a Lie algebra $\gotK$:
$$ M\cong V\cong \gotK ^* .$$
It follows that $\Ph M = \Ph \gotK ^* = T^* K$, where $K$ is the
group having $\gotK$ as its Lie algebra \cite{qcp}. The group
$K$ has an obvious semi--direct product structure. It is
convenient to realize this group as $\bR ^3\times \bR$ with the
group multiplication given by
\be\label{mno}
(\vec{p},p_0)({\vec{p} }',p_0')=
\left(\vec{p}\exp \left(\frac{h}{2} p_0'\right) +
\exp \left(-\frac{h}{2} p_0\right) {\vec{p}}{\,}', p_0
+p_0'\right),
\ee
because then $K$ is easily recognized as the quotient group of
$G^*$ described in \cite{luki}, corresponding to the momentum
coordinates $P_0$, $P_1$, $P_2$, $P_3$ (as it should be, to the
Poisson inclusion $\gotK ^*\subset G$ there corresponds dually
the Poisson projection $G^*\to K$).

According to the general rule, we should now compute the moment
map $J\colon \Ph M\to G^*$ implied by the Poisson action of $G$
on $M$. However, in order to calculate the mass shell in $\Ph m$
it is sufficient to compute only the translational momentum part
$\Ph M \to G^*\to K$ of $J$, because the first {\em Casimir}
\cite{Lukier,Lukier1} (valid also in the Poisson context),
\be
C_1 = \left(\frac{\sinh \frac{h}{2}p_0}{\frac{h}{2}}\right)^2 +
\sum_{k=1}^3 p_kp^k,\qquad\qquad p^k\equiv -p_k,
\ee
depends only on $p$'s (this standard form of $C_1$ holds in the
parametrization of $K$ as in (\ref{mno})). But the projected $J$
is nothing else but the moment map for the action restricted to
the subgroup $V\subset G$. Since the action of $V$ on
$M=V=\gotK ^*$ is just the addition, the corresponding moment
map is just the projection $T^*K\to K$ in the cotangent bundle.

We conclude that the mass shell in $\Ph M =T^*K$ for the mass
$m$ is given by
$$ \{ (p,x)\in T^*K :  C_1(p) = m^2\}.$$
Here $p\in K\cong \bR ^3\times \bR$ and $x$ denotes the dual
(canonically conjugate) variable (assume it takes values in $\bR
^3\times \bR $).

\section{Trajectories}

The equations of characteristics on the mass shell are
obtained simply by calculating the flow of $C_1$ in $\Ph M$:
$$ \dot{x}^k =\{ \frac12 C_1 , x^k\} = p^k,\qquad\qquad
\dot{x}^0 = \{ \frac12 C_1 , x^k\} = \frac{\sinh hp_0}{h},
\qquad\qquad k=1,2,3.$$
As shown in \cite{woj}, the velocity
$$ \frac{dx^k}{dx^0} = \frac{p^k}{\frac{\sinh hp_0}{h}} =
\frac{p^k}{\sqrt{p^2 +m^2}\sqrt{1 + (\frac{h}{2} )^2(p^2 + m^2)}}
$$
is not a monotonic function of the momentum. However, the
variables $x$ are here not the natural position coordinates. The
intrinsic coordinates implied by the structure of the twisted
cotangent bundle come from the left projection in $\Ph M$, which
in our case is just the right translation of a covector in
$T^*K$ to the group unit \cite{qcp}:
\be
q^k = x^k \exp \left(\frac{h}{2} p_0
\right),\qquad\qquad q^0 = x^0
-\frac{h}{2}\sum_{j=1}^3p_jx^j .
\ee
 One can easily check that these
coordinates have Poisson brackets exactly as in (\ref{pb}).
The flow of $C_1$ gives now
$$
 \dot{q}^k = p^k \exp
 \left(\frac{h}{2}p_0
 \right), \qquad \dot{q}^0 =
\frac{\sinh hp_0}{h} + \frac{h}{2}{\vec{p}\,}^2,
$$
hence the velocity in terms of these coordinates has the
following square
\be
\left(\frac{d\vec{q}}{dq^0}\right)^2 =
\frac{
\bigg(\exp (hp_0) -1\bigg)^2 - m^2 h^2 \exp
\left(hp_0
\right)}{
\left(\exp (hp_0) -1
-\dis \frac{m^2 h^2}{2}
\right)^2}.
\ee
Using the monotonic change of variables $t\equiv \exp (hp_0) -1$,
we obtain
\be
\left(\frac{d\vec{q}}{dq^0}\right)^2 = f(t) =
\frac{t^2 - m^2 h^2 (t+1)}{
\left(t-\dis\frac{m^2h^2}{2}
\right)^2}.
\ee
An elementary calculus shows that
\ben
\item $\dis{\lim_{t\to\infty} f(t) =1}$ \ \ (from below)
\item $\dis{f'(t_0)=0 \Longleftrightarrow t_0=
 \frac{m^4h^4/2}{4-m^2h^2}}$.
 \een
{}From these facts it follows that $f$ is a monotonic function,
with the limt at infinity equal 1.

\end{document}